\documentclass{solarphysics}

\usepackage[hyperref,optionalrh,showbiblabels]{spr-sola-addons} % For Solar 
\usepackage{epsfig}          % For eps figures, old commands
\usepackage{graphicx}        % For eps figures, newer & more powerfull
\usepackage{color}           % For color text: \color command
\usepackage{breakurl}        % For breaking URLs easily trough lines
\usepackage{natbib}

\newcommand{\etal}{{\it et al.}}

\chardef\us=`\_

\begin{document}

\begin{article}
\begin{opening}

\title{Non-neutralized electric currents in solar active regions and flare productivity}

%%%%%%%%%%%%%%%%%%%%%%%%%%%%%%%%%%%%%%%%%%%%%%%%%%%
%% Authors Names
%
% \author[addressref={},corref,email={}]{\inits{}\fnm{}\lnm{}\orcid{}}
\author{Ioannis~\surname{Kontogiannis}$^{1}$\sep
        Manolis~K.~\surname{Georgoulis}$^{1}$\sep
        Sung-Hong~\surname{Park}$^{2}$\sep
        Jordan~A.~\surname{Guerra}$^{2}$\sep
          }
%%%%%%%%%%%%%%%%%%%%%%%%%%%%%%%%%%%%%%%%%%%%%%%%%%%
%% Runningheads
%
\runningauthor{Kontogiannis \etal}
\runningtitle{Non-neutralized currents and flare productivity}

%%%%%%%%%%%%%%%%%%%%%%%%%%%%%%%%%%%%%%%%%%%%%%%%%%%
%% Affilations
%% id shold be the same with \author addressref value.
%\address[id={}]{}
  \institute{$^{1}$ Research Center for Astronomy and Applied Mathematics (RCAAM) Academy of Athens, 4 Soranou Efesiou Street, Athens, GR-11527, Greece
                     email: \url{jkonto@noa.gr} \\
             $^{2}$  School of Physics, Trinity College Dublin, Dublin 2, Ireland
                     email: \url{shpark@tcd.ie} email: \\}
%%%%%%%%%%%%%%%%%%%%%%%%%%%%%%%%%%%%%%%%%%%%%%%%%%%

\begin{abstract}
We explore the association of non-neutralized currents with solar flare occurrence in a sizable sample of observations, aiming to show the potential of such currents in solar flare prediction. We use the regularly produced high quality vector magnetograms by the Helioseismic Magnetic Imager and more specifically the Space weather HMI Active Region Patches (SHARP). Through a newly established method, that incorporates detailed error analysis, we calculate the non-neutralized currents contained in active regions (AR). Two predictors are produced, namely the total and the maximum unsigned non-neutralized current. Both are tested in AR time-series and a representative sample of point-in-time observations during the interval 2012-2016. The average values of non-neutralized currents in flaring active regions are by more than an order of magnitude higher than in non-flaring regions and correlate very well with the corresponding flare index. The temporal evolution of these parameters appears to be connected to physical processes, such as flux emergence and/or magnetic polarity inversion line formation that are associated with increased solar flare activity. Using Bayesian inference of flaring probabilities, it is shown that the total unsigned non-neutralized current outperforms significantly the total unsigned magnetic flux and other well established current-related predictors. Thus, it shows good prospects for inclusion in an operational flare forecasting service. We plan to use the new predictor in the framework of the FLARECAST project along with other highly performing predictors.
\end{abstract}
%%%%%%%%%%%%%%%%%%%%%%%%%%%%%%%%%%%%%%%%%%%%%%%%%%%
%% Keywords
%

\end{opening}
%-------------------------------------------------

\section{Introduction}
     \label{S-Introduction} 

Solar flares are localized and intense brightenings of the solar atmosphere, much brighter than the background and evident throughout the entire electromagnetic spectrum. They are associated with in situ acceleration of energetic particles and (often) coronal mass ejections (CMEs), comprising some of the most spectacular and energetic phenomena of the solar system \citep{fletcher11}. The overall associated energy, released in very short time, is even exceeding $10^{32}$\,erg in very large events encompassing both thermal and non-thermal processes.

The soft X-ray flux of a flare in the 1-8\,{\AA} range, as recorded by the Geostationary Operational Environmental Satellites (GOES), is used to categorize solar flares in order from largest to smallest in a logarithmic scale of classes X, M, C, B and A. These scales are complemented by decimal subclasses (e.g. M1.0, C5.2, etc). Flares of M1.0 and higher are often referred to as major flares and are the ones that mostly affect the geospatial environment. On the other hand, B- and A-class flares are the weakest and often lie within the soft X-ray background produced by the global solar atmosphere.

Flares and CMEs affect the geospatial environment in diverse ways, at timescales ranging from a few minutes to days. In flares, these effects are immediate. The electron density in a range of ionospheric altitudes is affected by the enhanced X-ray and EUV radiation, disrupting radio communications while the subsequent expansion of the atmosphere increases the drag on low-altitude satellites. Space-born instrumentation and crew are also vulnerable to direct flare-related electromagnetic radiation and particles. This evident absence of early warning for flares and flare-related effects has necessitated the pursuit for accurate flare prediction. The scientific community has intensified efforts in this direction by incorporating the constant flow of solar magnetic observations achieved during the last decades. This is done towards two complementary and often overlapping directions: on one hand, to understand the fundamental physics behind the flare phenomenon \citep{shibata11}. On the other hand, to develop efficient flare prediction schemes \citep[see e.g.][]{georgoulis12b}.

Energy released in flares is known to be stored in the intense, complex magnetic field configurations of solar active regions. The ability of magnetic fields to store energy is associated with their departure from a current-free (potential) state. This non-potentiality is evident in optical (H$\alpha$, H$\beta$), EUV and X-ray images of the active region solar corona, where significant twist of the coronal loops is seen \citep{Leka96,schrijver16}. This twist requires the presence of substantial amounts of field--aligned electric currents. Therefore, the magnetic field configuration and the electric current distribution offer two aspects of the same physical reality \citep[see e.g.][and references therein]{melrose95}. Since inferring the electric currents requires the ability to record all three components of the magnetic field vector, studies focusing on electric currents and their role in flares started after vector magnetograms became widely available \citep[see e.g.][and references therein]{canfield93,zhang95,Leka96}. Nowadays, the regular flow of high temporal and spatial resolution photospheric vector magnetograms has boosted research in this topic \citep[e.g.][]{ravindra11,gosain14,janvier14,vemareddy15,inoue15}.

A pertinent issue is whether electric currents in solar active regions are neutralized. In the ideal case of an isolated twisted flux tube embedded in a field-free medium, the net current along the tube should be zero, being equal to the sum of two opposite directed currents: a direct (volume) current along its axis that produces its twist and a return (surface) current that isolates the flux tubes from the field-free environment \citep{parker79}. Current neutralization subsequently means that direct and return currents balance each other within a given magnetic polarity of an active region, giving rise to a zero net current per polarity. However, early calculations based on vector magnetograms showed that high values of net currents exist within active regions, implying a non-neutralization situation and a subsequent injection of net currents in the solar corona \citep{melrose91,melrose95}.

Since then, a series of studies addressed the non-neutralization of photospheric electric currents and their origin. Most of them have shown that there are weak or no return currents in the photosphere \citep{Leka96,mcclymont97,semel98,wheatland00,falconer01} and it is now being largely accepted that the currents that run along the magnetic field lines are non-neutralized. However, there have been some opposing arguments, reporting the existence of return currents in isolated sunspots \citep{wilkinson92}. Also, according to \citet{parker96}, the observational limitations of magnetographs consequently lead to erroneous calculations of net electric currents with no physical meaning, when the differential form of Amp\'{e}re's law is used. 

To address the issues of spatial resolution and errors that may hinder the correct interpretation \citet{georgoulis12a}, proposed a detailed methodology for the calculation of electric current neutralization. They also incorporated a detailed error analysis and imposed strict criteria on current neutralization. Their study of two active regions showed that intense non-neutralized currents are found exclusively at the vicinity of strong magnetic polarity inversion lines (MPIL) and that higher values are linked with higher flare productivity. Strong MPILs separate opposite polarities and are characterized by high magnetic field strength, leading to a plasma $\beta$ of order unity at the photosphere. In such conditions, the Lorentz force can overcome the hydrodynamic inertia of the photospheric plasma and produce shear. It was, therefore, concluded that non-neutralized currents are injected in the atmosphere with the emergence of flux and the shear observed in MPILs is generated by the Lorentz force when the cylindrical symmetry of the flux-tube footprints breaks down. Their results have not been reproduced since on a larger sample but the exclusive relation between non-neutralized currents in MPILs and shearing/twisting motions has been demonstrated via observations and models in subsequent studies \citep{janvier14,torok14,vemareddy15,dalmasse15}. Although there is no general consensus on the causal relationship between shearing motions and the Lorentz force, there seems to be a consensus that substantial net currents are injected in the corona along strong MPILs. Dissipative processes (magnetic reconnection) are required for current neutralization, justifying the use of current--related quantities as flare predictors.

Electric current densities have shown good correlation with flaring and CME activity \citep{falconer02,yang12} and have been incorporated in flare prediction schemes in several studies \citep{leka_barnes03a,leka_barnes03b,leka_barnes07,bobra_couvidat}. In these studies, electric current related predictors rank among the best performing, motivating us to investigate new ones, based on the systematic approach of \citet{georgoulis12a}, where the effects of numerical artefacts and magnetic field measurement errors are mitigated and the contribution of MPILs with strong shear, as potential flaring sites in active regions, are intensified. Thus, we extend their results to a large sample of active regions, aiming to produce a new current-based predictor for use in the context of the FLARECAST project.

FLARECAST is a novel endeavour whose purpose is to produce real time flare predictions of unmatched accuracy. It aims to do so by incorporating the most efficient predictors and methods in a highly sophisticated prediction scheme, exploiting advanced machine-learning techniques. Along with utilizing predictors proposed so far in the literature, part of the project is devoted to developing new ones. In this context, we investigate the flare-prediction capability of the total and maximum non-neutralized electric currents in active regions. Our sample is statistically significant and consists of thousands of space-based, active-region vector magnetograms acquired during the present solar cycle 24. 

\section{Data set description: SHARP near real-time data}
\label{S-data}
We use observations from the Helioseismic Magnetic Imager \citep[HMI;][]{hmischerrer,hmischou} on board the Solar Dynamics Observatory mission \citep[SDO;][]{sdo}. Aiming to produce data specifically directed to space weather research, the HMI team have developed the Space weather HMI Active Region Patches \citep[SHARP;][]{bobra14}. These are cut-outs that contain the magnetic field vector components, remapped and deprojected as if observed at the solar disk center along with a set of quantities used for predicting solar eruptions. In this study we use the near-real time (NRT), cylindrical equal area (CEA) SHARP data. Aiming to use the non-neutralized currents as predictors in an automated, flare prediction service which will rely on the automatically produced HARP cut-outs, no constrains on the solar disk location have been imposed in this study.

To initially test the algorithm and its performance and whether there is a correspondence between the evolution of the parameters and the flaring activity of active regions, a sample of eleven National Oceanic and Atmospheric Administration (NOAA) active regions (AR) was chosen. These AR and the corresponding time-series durations, starts and ends were selected randomly and show none, low, moderate or high flare productivity. Table~\ref{Table:t1} summarizes the time-series start and end, the Mt.Wilson classification type as well as the corresponding flare productivity of the active regions.

For further testing, we processed a representative sample of SHARP cut-outs. 336 random days were selected between September 2012 and May 2016, which comprise the 25\,\% of the entire SHARP data coverage. For each day, we processed SHARP cut-outs at a cadence of 6\,h, resulting to 9454 data points.  

The database of the Geostationary Operational Environmental Satellite (GOES\footnote{\url{http://www.ngdc.noaa.gov/stp/satellite/goes/}}) is utilized to derive the flare association of the SHARP observations. For every HARP cut-out of our sample (both active region time--series and the representative sample), we derived the flare onset times and classes, within the following 24\,h. Specifically for the time series, we also calculated the total flare index for the duration of each time-series, using the formula of \citet{abramenko05}, that is:

\begin{equation}
FI=(100S^{(X)}+10S^{(M)}+1.0S^{(C)}+0.1S^{(B)})/\Delta t,
\label{equation:fi}
\end{equation}

\noindent where $S^{(X,M,C,B)}$ is the, per class, sum of flare magnitude (the number that follows the class letter) and $\Delta t$ is the time interval in days. The number of flares per class and the total flare index for the AR sample are shown in columns 4-8 of Table~\ref{Table:t1}. Concerning the representative sample of SHARP CEA NRT cut-outs, out of the 9454 points of the sample, 2573, 212 and 16 are associated with C-, M- and X-class flares,  respectively, over the preset window.

\section{Analysis: Non-neutralized currents in solar active regions}

We follow the method of \citet{georgoulis12a}, which incorporates measurement uncertainties and poses strict criteria on deciding whether the calculated currents are neutralized. Here we briefly describe the method, while the reader is referred to the original paper for more details.

The map of the radial component of the SHARP vector magnetic field is partitioned into non-overlapping patches of the same magnetic polarity, using a gradient--based, flux tessellation scheme \citep{barnes05}. This partitioning method has also been used by \citet{georgoulis07} for the calculation of the effective connected magnetic field strength. To ensure that quiet Sun magnetic elements will be excluded and only the sizable, active-region-only magnetic polarities will be taken into account, thresholds on magnetic field strength, enclosed magnetic flux and area are imposed. According to \citet{georgoulis12a}, small changes in the selected thresholds do not alter the essence of the results. Their conclusion was also verified in our dataset by testing thresholds, e.g. magnetic field strength up to 300\,G and enclosed magnetic flux up to $10^{20}$\,Mx. Using lower thresholds will result in more partitions, which will not contribute to the calculations but will dramatically slow up the process. The thresholds selected in magnetic field strength, enclosed magnetic flux and size per partition are 100\,G, 5$\cdot10^{19}$\,Mx and 40\,pixel correspondingly. Prior to partitioning, the input magnetograms were smoothed, by a $5\times5$ pixel kernel, to produce partitions with slightly smoother boundaries and (in the case of time--series) avoid dramatic changes of the partitioning process from frame to frame. The process returns all non-overlapping partitions, along with their flux weighted centroid positions, boundaries and corresponding magnetic fluxes.

For each partition, the vertical electric current density is calculated using the differential form of Amp\'{e}re's law:

\begin{equation}
J_{z}=\frac{1}{\mu_{0}}(\frac{\partial B_{y}}{\partial x}-\frac{\partial Bx}{\partial y})
\label{equation:ampere}
\end{equation}

\noindent where $B_{x}$, $B_{y}$ are the two horizontal components of the magnetic field and $\mu_{0}$ is the magnetic permeability of the vacuum. The total current $I_{i}$ contained in each partition is calculated as the algebraic sum of the contained $J_{z}$. The error maps of the three components of the magnetic fields are used to calculate the corresponding error $\delta I_{i}$ for each $I_{i}$. To decide whether $I_{i}$ is non-neutralized, we calculate the corresponding potential magnetic field components \citep{alissandrakis81} using the vertical magnetic field $B_{z}$ as boundary condition and repeat the calculation for the potential magnetic field configuration. The value of current $I_{i}^{pot}$ in this case should be, by definition, equal to zero. Due to numerical effects, however, non-zero results are produced. The total electric current, $I_{i}$, of a given partition is considered non-neutralized, $I_{i}^{NN}$, if it satisfies the following criteria: $I_{i} > 5\times I_{i}^{pot}$ and $I_{i} > 3\times \delta I_{i}$. From the set of non-neutralized partition currents (if any) within the active region at a given time, we calculate the maximum value, $I_{NN,max}$, and the total unsigned non-neutralized current $I_{NN,tot}$. The latter is defined as the sum of the absolute current values of the non-neutralized partitions, $I_{i}^{NN}$ i.e.:

\begin{equation}
I_{NN,tot}=\sum_{i}|I_{i}^{NN}|
\label{equation:nncur}
\end{equation}

\noindent The rationale for selecting these two quantities is, as already mentioned, that strong flares in ARs are associated with MPILs, which are in turn associated with strong non-neutralized currents \citep{georgoulis12a,dalmasse15,vemareddy15}. Several sites in an active region may develop suitable conditions for flaring. $I_{NN,max}$ may be an indicator of the non-neutralized current that has increased at a certain site while, if several such locations are indeed present (hence $I_{NN,tot}$ is increased), then the active region should be even more prone to strong flares.

It should be pointed out that our methodology ensures that the calculated parameters are not size dependent but exclusively linked to the presence of strong MPILs. In this sense, $I_{NN,tot}$ and $I_{NN,max}$ are not extensive but intensive and therefore, neither the size of an active region nor the extent of the cut-out (so long as it fully contains the active region) affect the calculations.  

We investigate the flare-predictive capability of $I_{NN,max}$ and $I_{NN,tot}$ in a twofold manner. First, we explore the time evolution of the non-neutralized currents of active regions in comparison with their flaring activity, in order to assess whether the results of \citet{georgoulis12a} extend to more than the two cases the authors presented. Then, the performance of the two parameters is examined on a sizable sample of point-in-time observations, which is representative of the sample that FLARECAST will rely on.

\section{Results}
\subsection{General trends, and rationale for using non-neutralized currents in the FLARECAST project}
\label{S-ar_sample}
The purpose of this section is to assess whether 1) more flare-productive AR are linked with higher non-neutralized currents, 2) the two quantities show an evolution similar to their flare productivity and 3) this evolution is apt for prediction, i.e. the increase of the two parameters precedes flares and shows some correlation with flare class or flare index within a given time-window.

\begin{figure}
 \centerline{\includegraphics[width=1\textwidth]{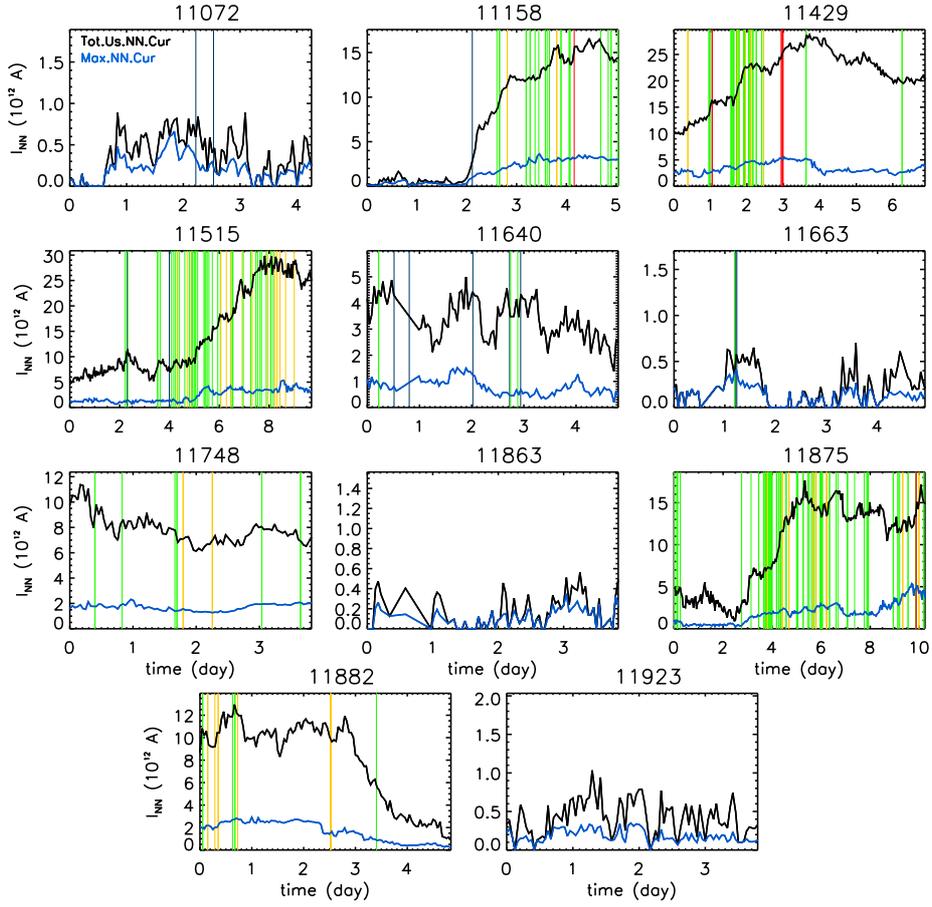}}
 \caption{$I_{NN,tot}$ (black) and $I_{NN,max}$ (blue) temporal evolution of the eleven ARs of Table~\ref{Table:t1}. Blue, green, yellow and red vertical lines mark the start times of B-, C-, M- and X-class flares.}
\label{fig:nncur_all_ar}
 \end{figure}

In Fig.~\ref{fig:nncur_all_ar} we plot the temporal evolution of the $I_{NN,tot}$ and $I_{NN,max}$ for each of the eleven active regions of the first sample. It is clear that flare productive AR have higher $I_{NN,tot}$ and $I_{NN,max}$ values, by more than an order of magnitude. The strongest flares require the existence of systematically high non-neutralized currents as e.g. in the case of AR\,11429, 11515, 11748 and 11875. Another interesting finding is that many strong flares are preceded by a few hours of increasing $I_{NN,tot}$ (e.g. AR\,11158, 11515, 11875). This increase corresponds to the emergence of new flux and to the development of a strong PIL and is also reflected on their magnetic type evolution (Table~\ref{Table:t1}, last column).

Active regions with $I_{NN,tot}$ and $I_{NN,max}$ lower than $10^{12}\,A$ exhibit very low or no flaring activity at all. For instance, AR\,11072 and AR\,11663 produced only B- or up to C1.1 class flares, respectively. In the latter case, these flares were clustered around relatively enhanced non-neutralized currents, during the second day. This clustering of flares during well defined peaks of $I_{NN,tot}$ is also observed in AR\,11640, which exhibits more frequent flaring activity, owing to its overall higher values of $I_{NN,tot}$ and $I_{NN,max}$.

To some extent, the more sizable, flare--productive active regions also exhibit repeated flaring activity during peaks of the two parameters, although these peaks do not stand out clearly but lie on an increased background. This is due to the fact that these active regions contain many non-neutralized partitions along intense MPIL, which increase the total amount of non-neutralized current. More careful inspection shows that even for these active regions, the long temporal increase of non-neutralized currents exhibits structure in finer temporal scales, manifested by shorter peaks on top of the general trend. This is seen mostly at the $I_{NN,tot}$ curve while the $I_{NN,max}$ curve is less inclined. This difference may be due to the fact that the non-neutralized current of a given partition may not increase above a certain limit. Instead, it is the increasing number of non-neutralized partitions that results in an overall increase of $I_{NN,tot}$. Note, also, that intense flaring activity is found at times when $I_{NN,tot}$ is several times higher than $I_{NN,max}$. We will return to this remark in the next section. By incorporating the added effect of multiple non-neutralized partitions within an active region, $I_{NN,tot}$ seems to represent its evolution more efficiently and, in this sense, it may prove a more efficient flare predictor than $I_{NN,max}$. 

Accumulation of non-neutralized currents in active-region partitions as part of an energy build-up process may be clearly seen in AR\,11158, 11429, 11515 and 11875. In AR\,11158, the flux emergence that is observed at the end of the second day is accompanied by an increase of the total non-neutralized currents, showing that flux emergence results in the injection of net currents in the active-region corona. This is in line with previous results on the origin of the non-neutralized currents. \citet{georgoulis12a} used a dimensionless non-neutralization factor to show that two active regions with very different flare productivity were coherent structures. This lead the authors to conclude that the non-neutralized currents they contained had a sub-photospheric origin. Their results were corroborated by the MHD simulations of \citet{torok14}, which showed that active regions are born with substantial non-neutralized currents. 

The association between non-neutralized current increase and flaring--CME activity has been implied in previous studies \citep[see e.g.][]{ravindra11,janvier14,vemareddy15}. In most of these studies, which prompt the evolution of an AR, this association was inferred a posteriori, by manually selecting regions of interest, based on the sites of recorded flaring activity (around MPILs). On the contrary, judging by Fig.~\ref{fig:nncur_all_ar}, it seems that the temporal variation of $I_{NN,tot}$ and $I_{NN,max}$ brings out the evolution of non-neutralized currents at the regions of interest as clear enhancements on top of the overall evolution. During the strongest flares, the total non-neutralized current continues to increase, showing that the free energy build-up process continues and flares remove only a part of this energy. According to \citet{janvier14}, a local increase of the current may be observed after the flare at the footpoints of the ribbons.

It has been shown that strong direct currents, with weak or no return currents (implying the existence of non-neutralized currents) are compatible with the existence of twisted flux ropes \citep{schrijver08}. For instance, AR\,11429 was one of the most flare productive of cycle 24, whose total non-neutralized current content is the largest of the sample (along with that of AR\,11515). AR\,11429 exhibited strong MPIL, with strong shear, indicative of strong deviations from non-potentiality, justifying the build-up of large amounts of $I_{NN,tot}$. It produced two X-class flares at the end of day 3, both accompanied by CMEs, which were preceded by the formation of magnetic flux ropes \citep{chintzoglou15,syntelis16}. The X-class flare of the well studied AR\,11158 has also been associated with a flux rope formation \citep{inoue15}. These comparisons are useful as they possibly link the proposed predictors with the underlying physics of the flaring phenomenon. A more detailed examination exceeds the scope of the present study and will be reserved for the future.

\begin{figure}
 \centerline{\includegraphics[width=1.\textwidth]{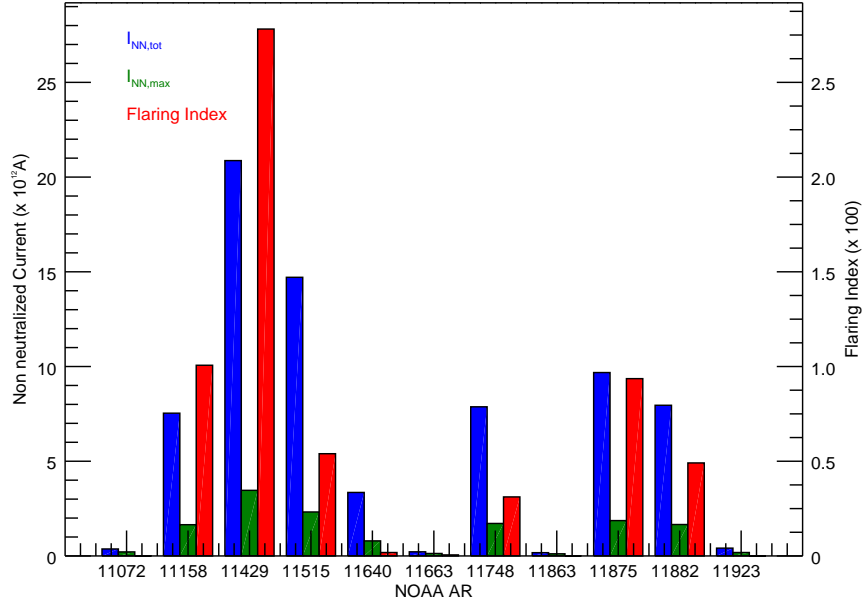}}
 \caption{Time averaged $I_{NN,tot}$ (blue) and $I_{NN,max}$ (green) for each active region of Table~\ref{Table:t1} and the corresponding FI (red), calculated for the entire duration of each time-series of Fig.~\ref{fig:nncur_all_ar}.}
\label{fig:ncur_fi_all_arr}
 \end{figure}

From a flare-forecasting point of view, one may coarsely infer $I_{NN,tot}$ thresholds of $0.5\cdot10^{12}$, $5\cdot10^{12}$ and $15\cdot10^{12}\,A$ for C-, M- and X-class flares to occur. However, apart from one case where a C-class flare occurred for $0.5\cdot10^{12}$\,A (AR\,11663), for the rest of the sample, a $I_{NN,tot}$ value of at least $3\cdot10^{12}$\,A was observed. Of course these thresholds are indicative: the actual merit of $I_{NN,tot}$ and $I_{NN,max}$ will be examined in the next section. Still, the good correspondence between $I_{NN,tot}$, $I_{NN,max}$ and flare productivity is also illustrated in Fig.~\ref{fig:ncur_fi_all_arr}. There, we plot the flare index during the observation interval for each AR of Fig.~\ref{fig:nncur_all_ar} along with the corresponding temporal averages of the two examined predictors. From this plot it is also confirmed that quiet and flare-productive ARs differ by more than an order of magnitude in terms of non-neutralized current content. The Pearson correlation coefficients between FI and $I_{NN,tot}$, $I_{NN,max}$ are 0.869631 and 0.865931, respectively. It is, therefore, concluded that the flare productivity of active regions is very well correlated with their amount of non-neutralized electric currents.

 \subsection{$I_{NN,tot}$ and $I_{NN,max}$ as solar flare predictors}
\label{S-nncur_predi}
Manually selected active region samples are useful to (visually) assess whether there is flare predictive potential in certain active region properties, in order to test the performance of the calculation algorithms and fine--tune the input parameters. Also, the time evolution of these properties often allows to derive some association with the physical processes involved in the flaring phenomenon. Nevertheless, these samples, however randomly selected, are inherently different than the ones actually used in automated prediction services, where photospheric magnetogram cut-outs are produced in real time and no (human biased) selection effects apply. To accommodate an examination of $I_{NN,tot}$ and $I_{NN,max}$ in ``realistic'' conditions, we use the representative sample of 9454 SHARP CEA NRT vector magnetograms.

\begin{figure}
  \centerline{
\includegraphics[width=0.8\textwidth]{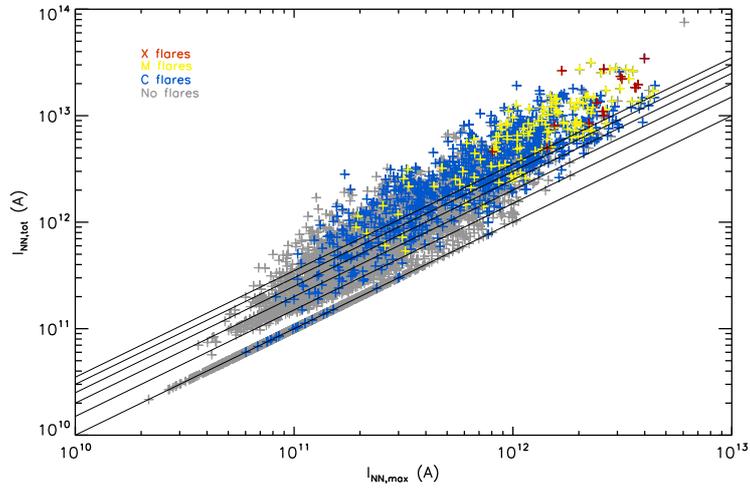}}
      \caption{$I_{NN,tot}$ vs. $I_{NN,max}$ (bottom) for the representative sample of SHARP CEA NRT cut-outs with non-zero non-neutralized currents. Lower to upper diagonal lines mark the $I_{NN,tot} = [1,1.5,2,2.5,3,3.5] \cdot I_{NN,max}$ fits.}
         \label{fig:nncur_imax}
\end{figure}

The corresponding non-zero $I_{NN,tot}$ and $I_{NN,max}$ are presented as a scatter plot in Fig.~\ref{fig:nncur_imax} along with color--coded flare association information. It is clearly seen that most cut-outs associated with major flares within the next 24\,h (marked with yellow and red crosses) are found towards the upper right corner i.e. are associated with higher non-neutralized electrical currents. Therefore, it seems that the two parameters allow a partial separation between flaring and non-flaring regions. For major flares to occur within the following 24\,h, coarse thresholds of $I_{NN,tot}$ and $I_{NN,max}$ may be derived: $5\cdot10^{11}\,A$ ($I_{NN,tot}$) and $1.9\cdot10^{11}\,A$  ($I_{NN,max}$) for M-class flares and $4.6\cdot10^{12}\,A$ ($I_{NN,tot}$) and $8\cdot10^{11}\,A$ ($I_{NN,max}$) for X-class flares. 

From the examination of the time evolution of the parameters at the previous section (Fig.~\ref{fig:nncur_all_ar}), it was found that intense flaring occurs at times when the difference between $I_{NN,tot}$ and $I_{NN,max}$ is largest while at quiet phases, the two quantities are roughly equal (or both close to zero). Fig.~\ref{fig:nncur_imax} allows us to extend these results to a larger sample. It can be seen that no major flares are found below the $I_{NN,tot} = I_{NN,max}$ line: M- and X-class flares require $I_{NN,tot} > 1.5 \cdot I_{NN,max}$ and $I_{NN,tot} > 3.5 \cdot I_{NN,max}$ respectively. Therefore, major flares require the presence of many non-neutralized partitions. A large number of non-neutralized partitions denotes the presence of a stronger, more fragmented MPIL. It is well established that strong complicated MPILs are linked with high non-potentiality of the magnetic field \citep[e.g.][]{falconer02,schrijver07}
and strong shear and are, consequently related to major flares. It is reasonable to suggest that if an AR contains several MPILS or one extended/more complicated MPIL, it is more prone to flare. While it seems that one single partition may reach a specific amount of non-neutralized current (reflecting to high $I_{NN,max}$), the presence of many non-neutralized partitions (either because there are several non-neutralized partitions along a MPIL or because there are several MPILS within an AR) results to $I_{NN,tot}$ being several times higher than $I_{NN,max}$. Therefore, the findings of Fig.~\ref{fig:nncur_imax} are an alternative way to demonstrate that strong, fragmented MPILs are required for major flares and that this requirement is effectively incorporated in the definition of $I_{NN,tot}$.

Next, we compare the performance of $I_{NN,tot}$ and $I_{NN,max}$ with that of the total unsigned flux, $\Phi_{tot}$, and the two electric current-related parameters given by the SHARP team, that is the total vertical current $J_{z,tot}$ and the sum of the modulus of the average net current per polarity $J_{z,sum}$ \citep{bobra14}. The two latter parameters express the amount of net current in an active region and the aim of the comparison is to investigate in practice whether $I_{NN,tot}$ and $I_{NN,max}$ have meaningful or trivial differences with these already established parameters. 

\begin{figure}    
 \centerline{\includegraphics[width=1.\textwidth]{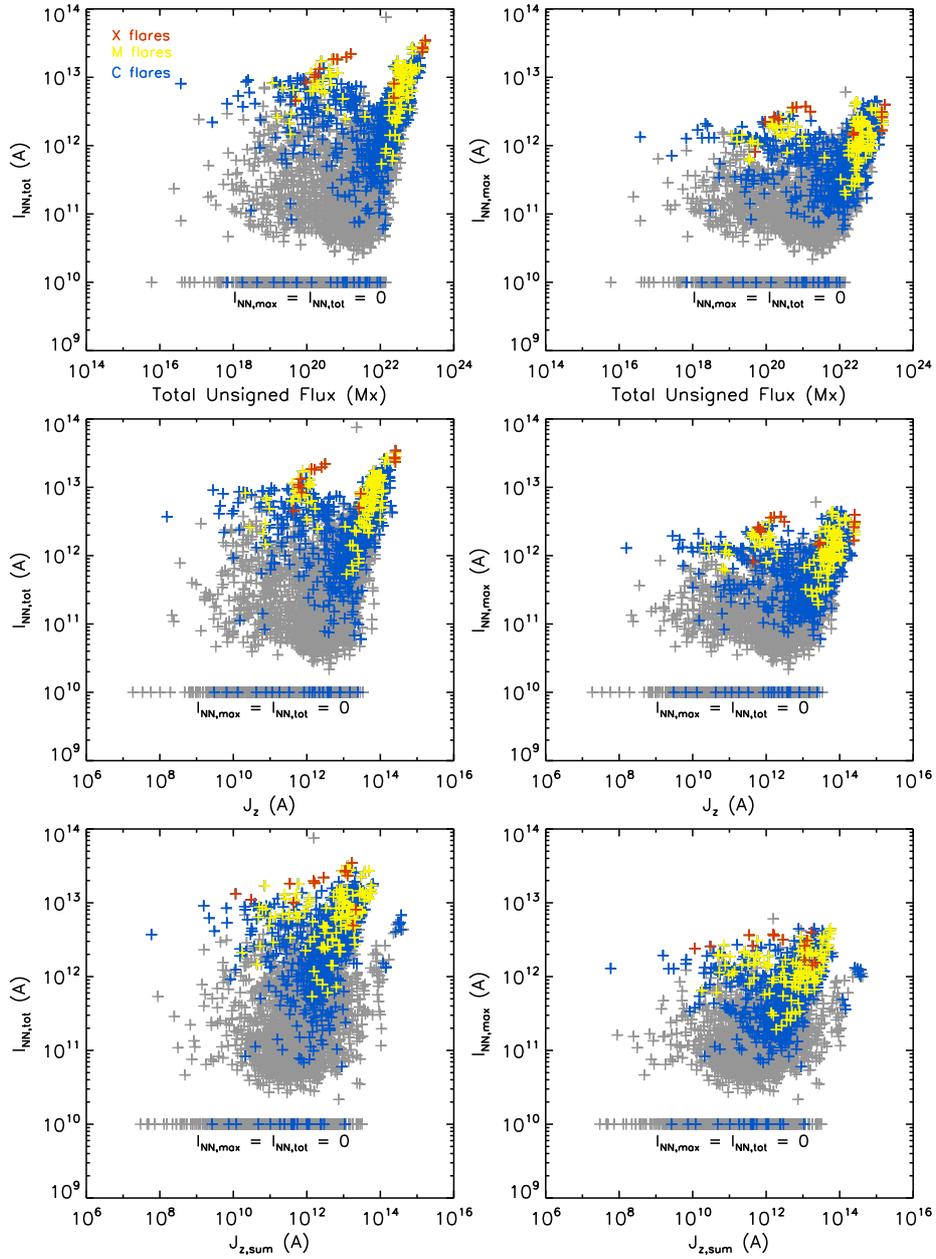}
 \vspace{0.07\textwidth}     }
 \caption{$I_{NN,tot}$ (left) and $I_{NN,max}$ (right) vs. $\Phi_{tot}$ (top), the total unsigned vertical current $J_{z}$ (middle) and the sum of the modulus of the average net current per polarity $J_{z,sum}$(bottom). The points associated with flares within 24\,h, are noted in different colors.}
\label{fig:nncur_flux}
 \end{figure}

The scatter plots between parameters calculated for each of the 9454 points are found in Fig.~\ref{fig:nncur_flux}. Cut-outs associated with flares of different classes that occur over the next 24 hours are denoted differently. The calculation process produced zero or no results for many cases. This is either because some cut-outs contained no non-neutralized partitions, or because the partitioning process was, at times, unable to provide eligible partitions as per the preset thresholds. Both cases may include decaying or developing AR without MPILs and are treated as corresponding to zero $I_{NN,tot}$ and $I_{NN,max}$. These points are included in all panels of Fig.~\ref{fig:nncur_flux} at ordinate equal to $10^{10}$\,A. 

In general, all panels in Fig.~\ref{fig:nncur_flux} imply, at best, a weak correlation between the examined parameters. The highest values of $\Phi_{tot}$ and $J_{z}$, correlate better with $I_{NN,tot}$ and $I_{NN,max}$. These points also correspond to increased flaring activity, as most of the major flares (yellow and red crosses) are concentrated there. 

Also, it is obvious that zero non-neutralized electric currents do not correspond to zero $\Phi_{tot}$, $J_{z}$ and $J_{z,sum}$. These are, by definition, extensive parameters, i.e. they increase with the size of the active region and/or the extent of the cut-out and acquire non-zero values. On the other hand, $I_{NN,tot}$ and $I_{NN,max}$ are intensive (non-extensive) parameters, as they depend only on the size and strength of the MPIL and are zero when no MPIL are formed, regardless of the cut-out extent. In our sample we find that only a very small fraction of these cut-outs ($\sim$0.5\%) are associated with C-class flares and none of them with M- or X-class flares.

\begin{figure}    %%%%%%%%%%%%%%%%%% FIGURE 2
                                % includes the two top panels
   \centerline{\hspace*{0.015\textwidth}
               \includegraphics[width=0.515\textwidth,clip=0]{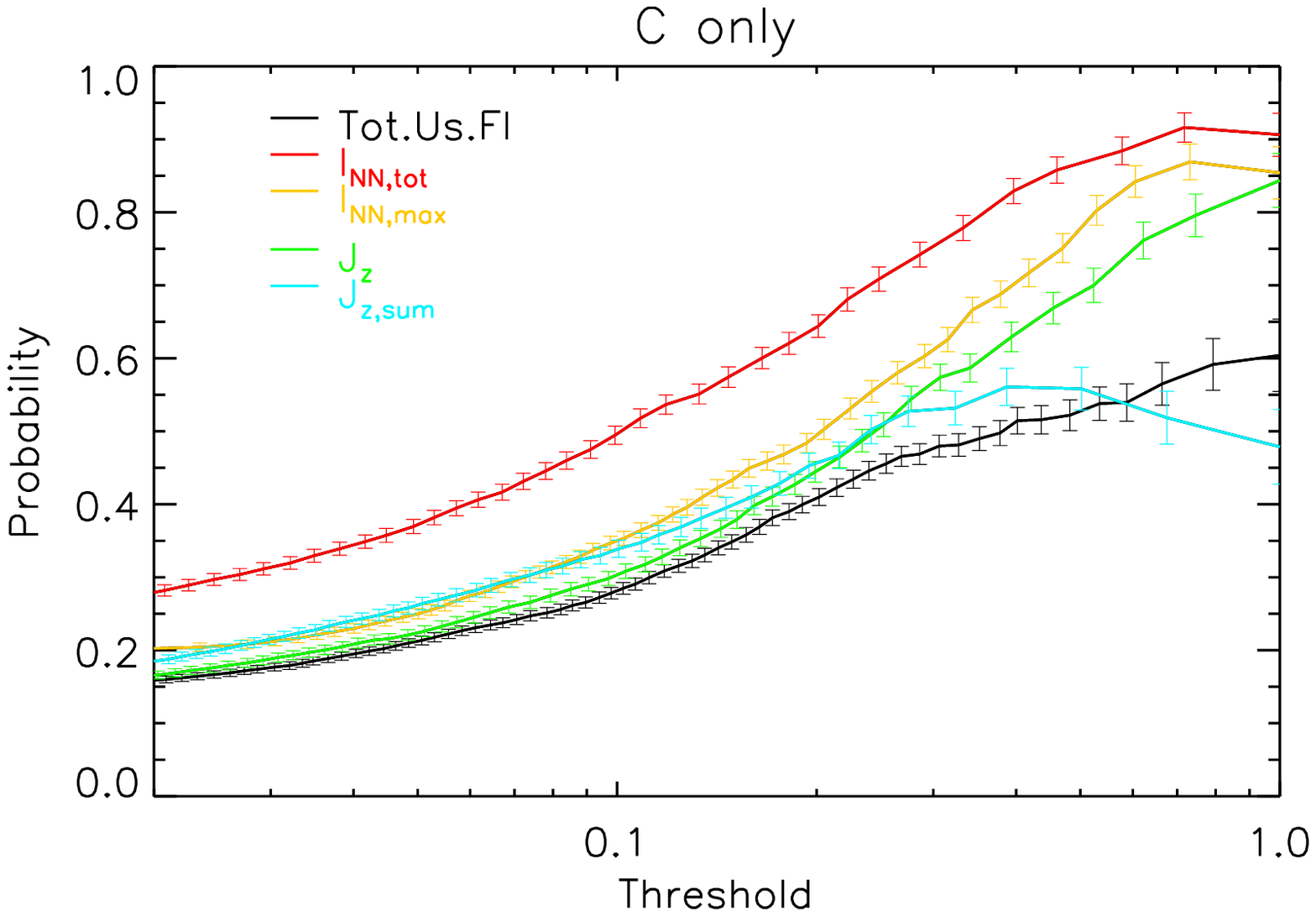}
               \hspace*{-0.03\textwidth}
               \includegraphics[width=0.515\textwidth,clip=0]{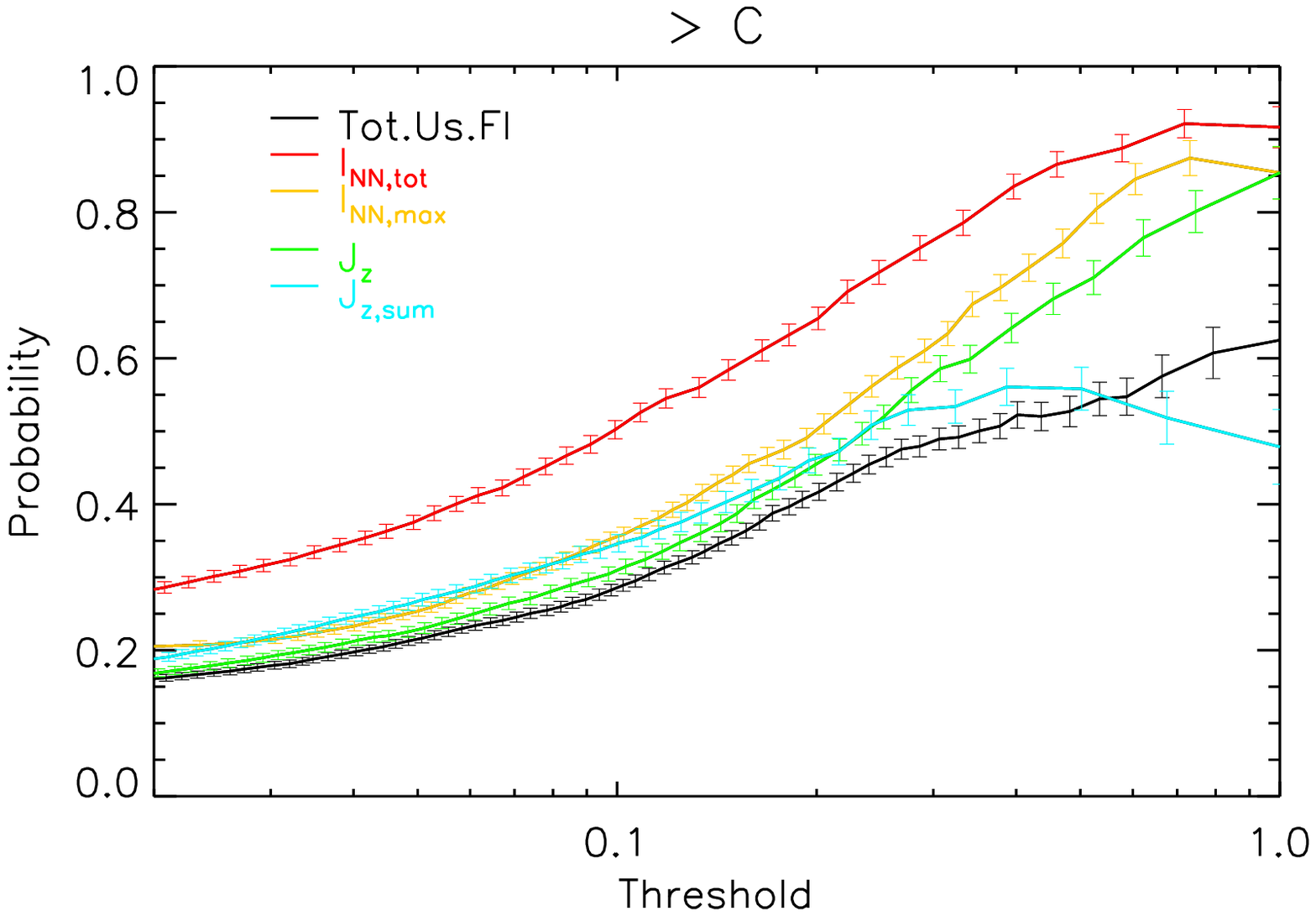}
              }
     \vspace{-0.35\textwidth}   % Shift close to the panel top
     \centerline{\Large \bf     % Includes the labels (here needs the color
                               %   package, see beginning of this file)
      \hspace{-0.1 \textwidth}  \color{white}{(a)}
     % \hspace{0.415\textwidth}  \color{white}{(b)}
        \hfill}
     \vspace{0.31\textwidth}    % Shift back to the panel bottom

   \centerline{\hspace*{0.015\textwidth}
               \includegraphics[width=0.515\textwidth,clip=0]{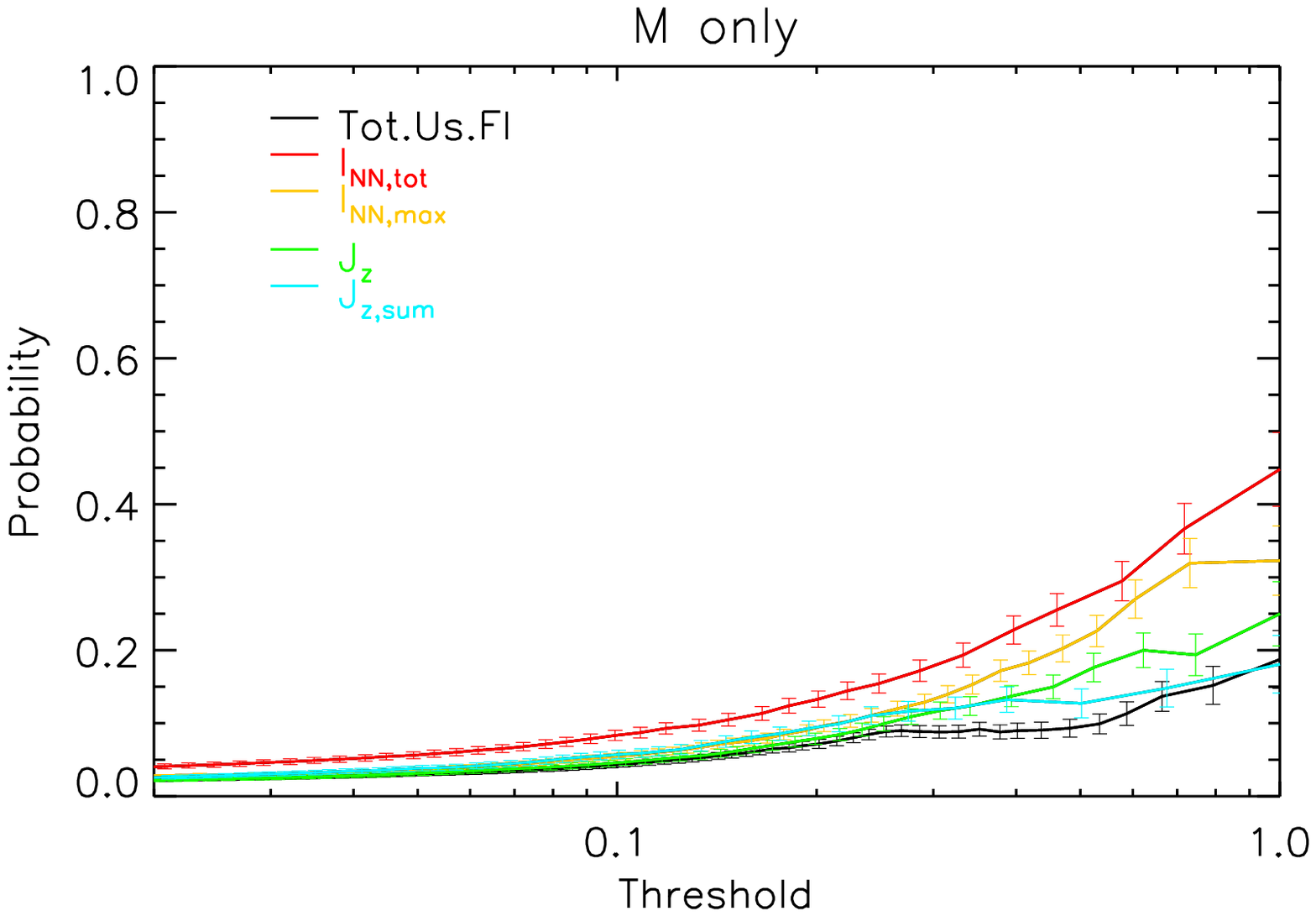}
               \hspace*{-0.03\textwidth}
               \includegraphics[width=0.515\textwidth,clip=0]{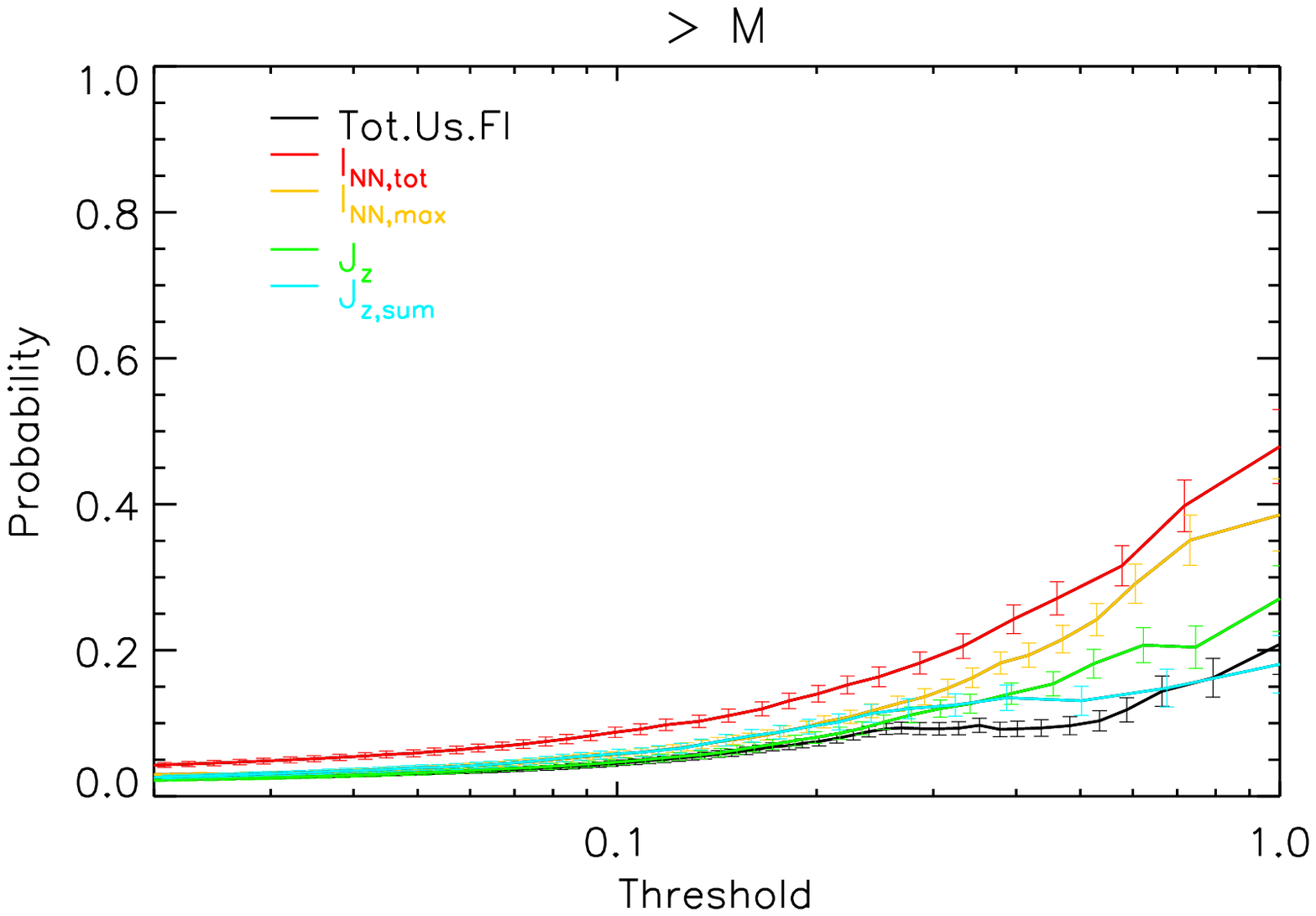}
              }
     \vspace{-0.35\textwidth}   % Shift close to the panel top
     \centerline{\Large \bf     % Includes the labels (here needs the color
                               %   package, see beginning of this file)
      \hspace{0.0 \textwidth}  \color{white}{(a)}
      \hspace{0.415\textwidth}  \color{white}{(b)}
        \hfill}
     \vspace{0.31\textwidth}    % Shift back to the panel bottom
   \centerline{\hspace*{0.015\textwidth}
               \includegraphics[width=0.515\textwidth,clip=0]{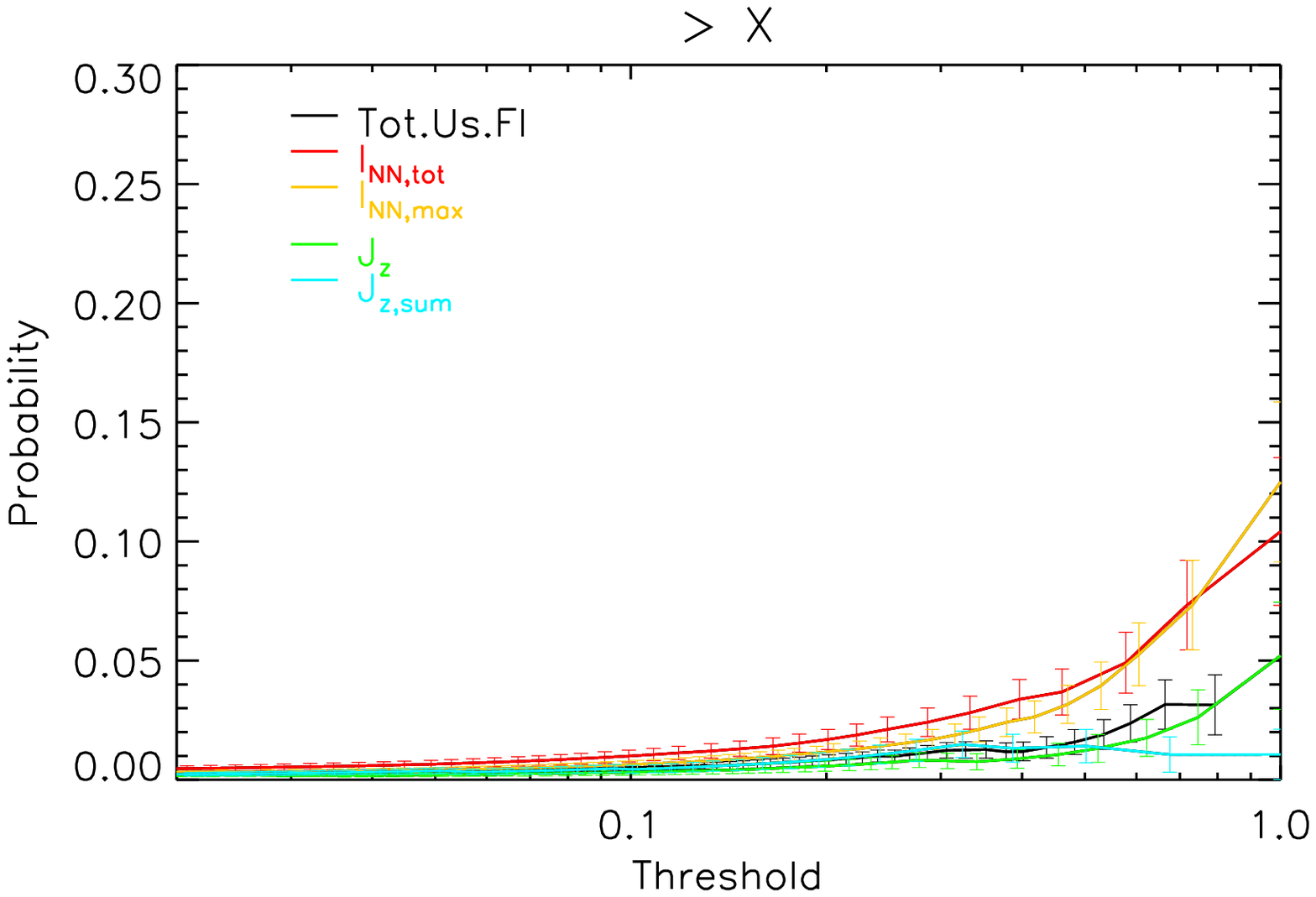}
               \hspace*{-0.03\textwidth}}

\vspace{-0.35\textwidth}   % Shift close to the panel top

\centerline{\Large \bf     % Includes the labels (here needs the color package)
      \hspace{0.0 \textwidth} \color{white}{(c)}
      \hspace{0.415\textwidth}  \color{white}{(d)}
         \hfill}
     \vspace{0.31\textwidth}    % Shift back to the panel bottom
 \caption{Bayesian inferred probabilities of $I_{NN,tot}$ (red) and $I_{NN,max}$ (orange) compared with that of $\Phi_{tot}$ (black),$J_{z}$ (green) and $J_{z,sum}$ (cyan), for C-, M-, X-, above C- and above M-class flares. $\Phi_{tot}$, $J_{z}$, $J_{z,sum}$, $I_{NN,tot}$ and $I_{NN,max}$ thresholds have been normalized to the corresponding maximum values, i.e. $1.68917\cdot10^{23}$\,Mx, $2.59441\cdot10^{14}$\,A, $4.21399\cdot10^{13}$\,A, $7.52898\cdot10^{13}$\,A and $6.06393\cdot10^{12}$\,A, respectively}
         \label{fig:flare_bayes}
 \end{figure}

The qualitative characteristics of the scatter plots of Fig.~\ref{fig:nncur_flux} are quantified, in terms of flare-forecasting potential, by the Bayesian inferred probabilities \citep{geo12,wheatland05} in Fig.~\ref{fig:flare_bayes}. For a given threshold, normalized to the maximum value of each parameter to facilitate comparison, we count the total number, N, of active regions with a parameter value higher than the threshold and the number, F, out of them, that are flaring within the next 24\,h. Then, the Bayesian probability is given by:

\begin{equation}
{p=\frac{F+1}{N+2}}
\label{bayes_prob}
\end{equation}

\noindent with uncertainty

\begin{equation}
{\delta p=\sqrt{\frac{p(1-p)}{N+3}}}
\label{bayes_error}
\end{equation}

The calculated Bayesian probabilities of the five parameters as functions of the normalized thresholds are shown in Fig.~\ref{fig:flare_bayes}. Both non-neutralized current parameters produce more significant probabilities than $\Phi_{tot}$, $J_{z}$ and $J_{z,sum}$ for a range of normalized thresholds. As expected, $I_{NN,tot}$ is more efficient. For C- and higher than C-class flares the difference with $\Phi_{tot}$ starts at $\sim$0.1 for low thresholds and surpasses 0.4. For major solar flares, the performance of $I_{NN,tot}$ is significantly better than the $\Phi_{tot}$. It should be noted that for very high thresholds, uncertainties increase, as the number of eligible data points (N) becomes smaller, decreasing the significance of the statistical sample. The results of Fig.~\ref{fig:flare_bayes} suggest that $I_{NN,tot}$ and $I_{NN,max}$ are more efficient than $\Phi_{tot}$, $J_{z}$ and $J_{z,sum}$ at segregating flaring from non-flaring active regions. They are, therefore, worth considering as flare predictors. 
\section{Conclusions and Discussion}
\label{s:conclusions}

In the context of an operational flare-forecasting service, we processed a sizeable sample of SDO/HMI SHARP cut-outs, aiming to examine the forecasting potential of two predictors related to the amount of non-neutralized electric currents contained in active regions. For the first time, we studied the evolution of the total and maximum unsigned non-neutralized currents, $I_{NN,tot}$ and $I_{NN,max}$, for eleven active regions, as well as their performance in flare prediction in a representative sample of SHARP CEA NRT data, consisting of 9454 cut-outs.

A brief discussion of the temporal evolution of these two new parameters showed that they correspond to physical processes such as MPIL formation and, as per previous studies, of a possible flux rope formation, closely related with intense flaring activity. Having focused on the potential of the $I_{NN,tot}$ and $I_{NN,max}$ as solar flare predictors, however, we did not discuss thoroughly the origin and evolution of non-neutralized currents in active regions. This is a long standing question that has been addressed through simulations and observations \citep{geo17} and a more detailed investigation is pending.

Both parameters are closely associated with flare productivity, confirming and extending the conclusions drawn by \citet{georgoulis12a}. It was shown for the first time, using a large representative sample of solar cycle 24, that both predictors are worth considering in automated flare prediction schemes. Out of the two proposed predictors, $I_{NN,tot}$ characterizes the flaring potential of an active region more efficiently because it encompasses the contributions of all possible flare-producing sites within an active region.

An important advantage of these predictors is that they include a detailed error analysis in the calculations of the non-neutralized currents. This approach has proven to be safe in concluding whether the currents in ARs are neutralized \citep{georgoulis12a}. The immediate comparison with the potential field ensures that the measured currents are existing and significant as per existing observations, not due to the numerical effects that arise when analytical expressions (such as Amp\'{e}re's law) are applied to discretized distributions of continuous physical quantities (such as the magnetic flux density). Therefore, possibly erroneous electric currents are constrained. Qualitatively, our results also suggest that the method used does not depend on the resolution of the processed magnetograms. The method was originally developed for Hinode SOT/SP magnetograms, the highest resolution available up to date, and in this study it was applied to the lower resolution SDO/HMI vector magnetograms. For all the reasons mentioned, $I_{NN,tot}$ is more advantageous than other commonly used current parameters, such as the total vertical current density, $J_{z}$, or the total unsigned current per polarity.

Thus said, the association between $I_{NN,tot}$ and $J_{z}$ is not trivial since by definition, there is no linear relationship between the two predictors. The total unsigned current of an active region is an extensive parameter, meaning that, in principle, it increases with the size of the active region (similarly to $\Phi_{tot}$). Non neutralized currents, on the other hand, should be zero for active regions that do not contain strong MPILs or for SHARP cut-outs that lack one polarity, regardless of their magnetic flux content and vertical electric current density. Therefore, $I_{NN,tot}$ and $I_{NN,max}$ are non-extensive parameters (as the lack of correlation in Fig.~\ref{fig:nncur_flux} also demonstrates), which incorporate in a physical way morphological information in active regions with possible (and plausible) implications for flare-triggering (i.e. strong and sheared MPILs). Given that current--related parameters have shown good prospects for flare prediction \citep{bobra_couvidat}, we expect improved results for $I_{NN,tot}$. Also, non-extensive parameters were found to outperform the extensive ones in CME prediction \citep{bobra16} which means that there may be some potential in extending the usage of $I_{NN,tot}$ to CME prediction as well.

In view of numerous studies on solar flare prediction, it has now become apparent that no single flare predictor suffices  to categorically forecast solar flares and that active regions are complex dynamical systems where more than one parameters are involved in flare triggering. Investigating the evolution and absolute magnitudes of non-neutralized electric currents is more attractive and tractable in the era of systematically and regularly available vector magnetogram. It is thus tempting to explore the possibility of producing, in the future, several predictors that involve such currents. For the moment, we are planning to include $I_{NN,tot}$ in the set of predictors that will feed machine learning algorithms (Florios et al. 2017, in prep.), in the framework of the FLARECAST project, at the same time investigating their physical significance via the project's explorative research component. 

%%%%%%%%%%%%%%%%%%%%%%%%%%%%%%%%%%%%%%%%%%%%%%%%%%%%%%%%%%%%%%%%%%%%%%%%%%%
 \begin{acknowledgements}
\noindent This research has been funded by the European Union's Horizon2020 research and innovation programme Flare Likelihood And Region Eruption foreCASTing" (FLARECAST) project, under grant agreement No.640216. The data used are courtesy of NASA/SDO, the HMI science team and the Geostationary Satellite
System (GOES) team. This work also used data provided by the MEDOC data and operations centre (CNES / CNRS / Univ. Paris-Sud), \url{http://medoc.ias.u-psud.fr/}.
\end{acknowledgements}

%\acknowledgment US spelling: \verb+\acknowledgment+
%\acknowledgement British  spelling: \verb+\acknowledgement+

%%%%%%%%%%%%%%%%%%%%%%%%%%%%%%%%%%%%%%%%%%%%%%%%%%%%%%%%%%%%%%%%%%%%%%%%%%%
 \appendix
\section{Differential versus integral form of Amp\'{e}re's law}

\begin{figure}[!ht]
   \centering
   \includegraphics[width=8cm]{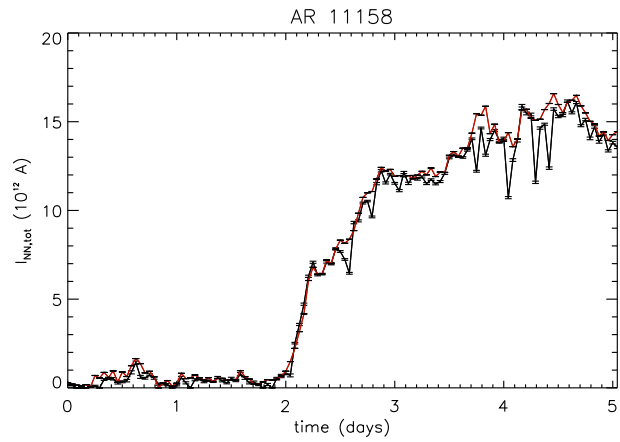}
      \caption{The total non-neutralized current, $I_{NN,tot}$ of AR\,11158 as a function of time, calculated using the integral (black) and differential (red) form of Amp\'{e}re's law.}
         \label{fig:diff_int}
\end{figure}

In the original work of \citet{georgoulis12a}, the integral form of Amp\'{e}re's law was chosen as being more accurate, albeit more time-consuming. To examine the trade-of between speed and accuracy we used both versions to calculate the total unsigned non-neutralized current for AR\,11158. The results are shown in Fig.~\ref{fig:diff_int}, where it is seen that the two versions produce practically the same results. For the longer part of the time-series, currents calculated by the two versions are the same within uncertainties. For most of the points, the differences are within the error and the two curves differ only at a few points. Overall, the differential form of Amp\'{e}re's law produces slightly smoother curves, since it utilizes a larger number of pixels, smoothing out differences from image to image (which could be due to noise or errors in the azimuth disambiguation and deprojection). Being based on the definition of a ``correct'' contour around each partition, the integral form is more sensitive to such variations. In view of this result, and given the fact that the integral form of Amp\'{e}re's law is significantly more time-consuming, we chose to use the differential form to calculate the electric currents that correspond to each partition.

%%% BIBLIOGRAPHY %%%%%%%%%%%%%%%%%%%%%%%%%%%%%%%%%%%%%%%%%%%%%%%%%%%%%%%%%%%

\bibliographystyle{spr-mp-sola}
     % name your Bibtex file containing your references (.bib)
\bibliography{references}

\begin{table}
\caption{NOAA AR sample. $t_{start}$,$t_{end}$ are the starting and ending dates of each time-series. B,C,M and X columns denote the number of the corresponding class flares within this interval and FI is the corresponding flare index while Mag.Type is the Mt.Wilson classification type of the NOAA AR.}
\setlength{\tabcolsep}{4.3pt}
\begin{tabular}{lcccccccc}
\hline
NOAA AR&$t_{start}$\,(UT)&$t_{end}$\,(UT)&B&C&M&X&FI&Mag.Type\\
\hline
11072&2010-05-20 16:24&2010-05-24 22:36&     2&       0&       0&       0&       0.06&$\beta$\\
11158&2011-02-10 21:59&2011-02-15 22:59&     1&      25&       4&       1&     100.67&$\beta$-$\beta\gamma$\\
11429&2012-03-04 01:23&2012-03-10 22:35&     0&      34&      12&       6&     278.15&$\beta\gamma$--$\beta\gamma\delta$\\
11515&2012-06-28 03:00&2012-07-07 19:48&     2&      39&      14&       0&      53.97&$\beta$--$\beta\gamma\delta$\\
11640&2013-01-01 03:35&2013-01-05 22:59&     5&       4&       0&       0&       1.81&$\beta$--$\beta\gamma\delta$\\
11663&2013-01-29 23:59&2013-02-03 22:23&     2&       2&       0&       0&       0.55&$\beta$\\
11748&2013-05-15 02:36&2013-05-18 22:24&     0&      10&       4&       0&      31.16&$\beta\gamma\delta$\\
11863&2013-10-10 02:35&2013-10-13 22:35&     0&       0&       0&       0&       0.0&$\alpha$--$\beta$\\
11875&2013-10-18 04:23&2013-10-28 10:23&     0&      81&      18&       2&      93.60&$\beta$--$\beta\gamma\delta$\\
11882&2013-10-26 02:23&2013-10-30 22:35&     0&       7&      10&       0&      49.10&$\beta\gamma\delta$--$\beta\gamma$\\
11923&2013-12-12 03:35&2013-12-15 22:35&     0&       0&       0&       0&       0.0&$\beta$\\
\hline
\end{tabular}
%\tablefoot{}
\label{Table:t1}
\end{table}

\end{article} 

\end{document}